\documentstyle[twocolumn,aps,amstex,amssymb,eqsecnum,epsf,psfrag]{revtex}

\begin{document}

\title{Well-Posed Initial-Boundary Evolution in General Relativity}

\author{
	B\'{e}la Szil\'{a}gyi${}^{1}$ ,
	Jeffrey Winicour${}^{1,2}$
       }
\address{
${}^{1}$ Department of Physics and Astronomy \\
         University of Pittsburgh, Pittsburgh, PA 15260, USA \\
${}^{2}$ Max-Planck-Institut f\" ur
         Gravitationsphysik, Albert-Einstein-Institut, \\
	 14476 Golm, Germany
	 }
\maketitle

\begin{abstract}

Maximally dissipative boundary conditions are applied to the initial-boundary
value problem for Einstein's equations in harmonic coordinates to show that it
is well-posed for homogeneous boundary data and for boundary data that is small
in a linearized sense. The method is implemented as a nonlinear evolution code
which satisfies convergence tests in the nonlinear regime and is stable in the
weak field regime. A linearized version has been stably matched to a
characteristic code to compute the gravitational waveform radiated to infinity.

\end{abstract}


\vspace{-1cm}

The waveform emitted in the inspiral and merger of a relativistic binary is
theoretical input crucial to the success of the fledgling gravitational wave
observatories. A computational approach is necessary to treat the highly
nonlinear regime of a black hole or neutron star collision. Developing this
computational ability has been the objective of the Binary Black Hole (BBH)
Grand Challenge \cite{gc} and other world wide efforts. The Grand Challenge
based a Cauchy evolution code on the Arnowitt-Deser-Misner (ADM)~\cite{adm}
formulation of Einstein's equations. Exponentially growing instabilities
encountered with that code have been traced to improper boundary
conditions~\cite{cboun}. (Even in the absence of boundaries, an ADM system
linearized off a Minkowski metric has a power law instability~\cite{cas}; this
triggers an exponential instability when the background Minkowski metric is
treated in non-Cartesian, e.g. spherical, coordinates.)  Other groups have
encountered difficulties in treating boundaries (see~\cite{lehn} for a recent
discussion) and the working practice is to forestall problems by placing the
outer boundary far from the region of physical interest (see e.g. ~\cite{mars})
or compactify the spacetime (see e.g.~\cite{husa}).

\begin{figure}
\psfrag{DD1}{${\cal D}_1$}
\psfrag{DD2}{${\cal D}_2$}
\psfrag{tt}{$t$}
\psfrag{SS}{${\cal S}$}
\psfrag{BB}{${\cal B}$}
\centerline{\epsfxsize=1.5in\epsfbox{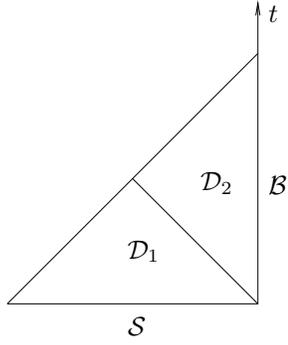}}
\vspace{0.5cm}
\caption{Schematic representation of the domain of dependence
${\cal D}_1$ of the initial value problem and the domain of
dependence ${\cal D}_1 \cup  {\cal D}_2$ of the IBVP.}
\label{fig:domain}
\end{figure}

This deficiency extends beyond numerical relativity to a lack of analytic
understanding of the initial-boundary value problem (IBVP) for general
relativity. The local-version of the IBVP is schematically represented in
Fig.~\ref{fig:domain}. Given Cauchy data on a spacelike hypersurface ${\cal S}$
and boundary data on a timelike hypersurface ${\cal B}$, the problem is to
determine a solution in the appropriate domain of dependence. Whereas there is
considerable mathematical understanding of the gravitational initial value
problem (for reviews see~\cite{crev}), until recently the IBVP has received
little attention (see e.g.~\cite{stewart,Friedrich98}). Indeed, only relatively
recently has the  method of maximally dissipative boundary conditions
~\cite{friedrichs,lax} been extended to the nonlinear IBVP with boundaries
containing characteristics~\cite{rauch,secchi2}, such as occurs in symmetric
hyperbolic formulations of general relativity. Friedrich and Nagy 
\cite{Friedrich98} have applied these methods to give the first demonstration
of a well-posed IBVP for Einstein's equations. The Friedrich-Nagy work is of
seminal importance for introducing the maximally dissipative technique into
general relativity. Their formulation, which uses an orthonormal tetrad, the
connection and the curvature tensor as evolution variables, is quite different
from the metric formulations implemented in current codes designed to tackle
the BBH problem. Although it is not apparent how to apply the details of the
Friedrich-Nagy work to other formalisms, the general principles can be carried
over provided Einstein's equations are formulated in the symmetric hyperbolic
form
\begin{equation}
  \sum_\alpha A^\alpha (u) \partial_\alpha u = S(u) 
   \label{eq:sh}
\end{equation}   
with coordinates $x^\alpha=(t,x,y,z)=(t,x^i)$ and evolution variables
$u=(u_1,...,u_N)$, where $A^\alpha$ are $N\times N$ symmetric matrices and
$A^t$ is positive-definite.

The simplest symmetric hyperbolic version of Einstein's equations employs
harmonic coordinates satisfying $H^\alpha:=\sqrt{-g}\Box x^\alpha
=\partial_\beta (\sqrt{-g}g^{\alpha\beta})=0$, in which the well-posedness of
the initial value problem was first established~\cite{bruhat,fisher}.
Well-posedness expresses the existence of a unique solution with continuous
dependence on the data. In the nonlinear case, existence is only guaranteed for
a short time, reflecting the possibility of singularity formation. Here we
show how this approach (i) can be applied to the IBVP problem in harmonic
coordinates, (ii) can be implemented as a robustly stable, convergent
3-dimensional nonlinear Cauchy evolution code and (iii) can be accurately
matched, in the linearized approximation, to an exterior characteristic
evolution code to provide the proper physical boundary condition for computing
the waveform radiated to infinity by an isolated source. Reference~\cite{Garf}
discusses the suitability of harmonic coordinates for numerical work
and for simulating the approach to a curvature singularity.

We base the evolution on the metric density
$\gamma^{\alpha\beta}=\sqrt{-g}g^{\alpha\beta}$, with $g =
\det(\gamma^{\alpha\beta})=\det(g_{\alpha\beta})$. As in Eq.~(B.87) of
Fock~\cite{fock}, we split the Einstein tensor into $G^{\alpha\beta}={\cal
E}^{\alpha\beta}+\frac{1}{2}g^{\alpha\beta}B-B^{\alpha\beta}$, where 
$B^{\alpha\beta}=-(-g)^{-1/2}\nabla^{(\alpha}H^{\beta)}$ vanishes when
$H^\alpha=0$, where
\begin{equation}
     {\cal E}^{\alpha\beta} =\frac{1}{2g}
         \gamma^{\mu\nu}\partial_\mu \partial_\nu
         \gamma^{\alpha\beta}  + S^{\alpha\beta}
\end{equation}
and where $S^{\alpha\beta}$ contains no second derivatives of the metric.
When the harmonic conditions $H^\alpha=0$ are satisfied, the reduced Einstein
equations ${\cal E}^{\alpha\beta}=0$ have principal part which is
governed by the wave operator $\gamma^{\mu\nu}\partial_\mu \partial_\nu$. In
terms of the evolution variables $u=(\gamma^{\alpha\beta}, {\cal
T^{\alpha\beta}, {\cal X}^{\alpha\beta}},{\cal Y}^{\alpha\beta}, {\cal
Z}^{\alpha\beta})$, where   ${\cal T^{\alpha\beta}}=\partial_t
\gamma^{\alpha\beta}$,   ${\cal X^{\alpha\beta}}=\partial_x
\gamma^{\alpha\beta}$,   ${\cal Y^{\alpha\beta}}=\partial_y
\gamma^{\alpha\beta}$ and  ${\cal Z^{\alpha\beta}}=\partial_z
\gamma^{\alpha\beta}$, we put these wave equations in symmetric hyperbolic
form (\ref{eq:sh}) by a standard construction~\cite{cour}.

We adapt the harmonic coordinates to the boundary so that the evolution region
lies in $z<0$, with the boundary fixed at $z=0$ in the  numerical grid. We
write $x^\alpha=(x^a,z)=(t,x,y,z)$ to denote coordinates adapted to the
boundary, so that $x^\alpha=(x^a,z)=(t,x^i)$ depending whether the Latin index
is near the beginning or end of the alphabet. We further adapt the coordinates
so that $\gamma^{za}|_{\cal B}=0$ (and hence ${\cal T}^{za}|_{\cal B}=0$). For
any timelike ${\cal B}$, these harmonic {\it gauge} conditions can be satisfied
and they are assumed throughout the following discussion.

We base the well-posedness of the homogeneous IBVP on Theorems 2.1 and 2.2 of
Secchi~\cite{secchi2}, which require that $u$ satisfy a boundary condition of
the form $Mu=0$, where $M$ is a matrix independent of $u$ and of maximal rank
such that the normal flux ${\cal F}^z=(u,A^z u)$ associated with an energy norm
be non-negative. Secchi's theorems include the present case where the boundary
is ``characteristic with constant rank'', i.e. $A^z$ has a fixed number of 0
eigenvalues.  The above  gauge conditions considerably simplify this maximally
dissipative condition for the reduced system ${\cal E}^{\alpha\beta}=0$. The
boundary matrix takes the form $A^z=\gamma^{zz} C$, where $C$ is a constant
matrix, and the flux inequality reduces to
\begin{equation}
      {\cal F}^z = -\sum_{\alpha ,\beta} {\cal T^{\alpha\beta}}
      {\cal Z^{\alpha\beta}} \ge 0.
\label{eq:maxd}      
\end{equation} Here ${\cal F}^z$ is identical to the standard energy flux for
the sum of 10 independent scalar fields.  This requirement can be satisfied in
many ways, e.g. by combinations of the homogeneous Dirichlet boundary condition
$\partial_t \gamma^{\alpha\beta}={\cal T^{\alpha\beta}}=0$, the homogeneous
Neumann condition $\partial_z \gamma^{\alpha\beta} ={\cal Z^{\alpha\beta}}=0$
and the homogeneous Sommerfeld condition $(\partial_t+\partial_z)
\gamma^{\alpha\beta} ={\cal T^{\alpha\beta}} +{\cal Z^{\alpha\beta}}=0$ on the
various field components. All these boundary conditions have the required form
$Mu=0$. The maximality of the rank of $M$ ensures that boundary conditions only
be applied to variables propagating along characteristics entering the
evolution region from the exterior \cite{kreiss}. For instance, assignment of a
boundary condition to the variable ${\cal T^{\alpha\beta}}-{\cal
Z^{\alpha\beta}}$, which propagates from the interior toward the boundary,
would violate (\ref{eq:maxd}).

Whereas the well-posedness of the IBVP for the reduced system can be
accomplished by a variety of boundary conditions, it can only be established
for the full system in a limited sense. The Bianchi identities and reduced
equations imply $\nabla_\mu (B^{\mu\nu}-\frac{1}{2}g^{\mu\nu}B)=0$, which has
the explicit form
\begin{eqnarray}
    \gamma^{\mu\nu}\partial_\mu \partial_\nu H^\alpha
      +C^{\alpha\mu}_\nu \partial_\mu H^\nu + D^{\alpha}_\nu H^\nu=0,
\label{eq:constrpr}
\end{eqnarray}
where $C^{\alpha\mu}_\nu$ and $D^{\alpha}_\nu$ depend algebraically on $u$ and
$H^\alpha$. Thus $H^\alpha$ obeys a symmetric  hyperbolic equation of the form
(\ref{eq:sh}). Harmonic Cauchy data satisfying the Hamiltonian and momentum
constraints and $H^\alpha =0$ also satisfy $\partial_t H^\alpha =0$ on ${\cal
S}$ by virtue of the reduced equations, so that uniqueness guarantees
$H^\alpha=0$ in the domain of dependence ${\cal D}_1$  (see
Fig.~\ref{fig:domain}) and hence the well-posedness of the Cauchy problem for
the full system. To extend well-posedness to the homogeneous IBVP, i.e. to
include region ${\cal D}_2$, we impose boundary conditions for the reduced
system that imply the homogeneous boundary conditions $H^z=0$ and $\partial_z
H^a=0$ for the harmonic constraints. Combined with the gauge condition
$\gamma^{za}=0$, the condition $H^z:=\partial_b \gamma^{zb}+\partial_z
\gamma^{zz}=0$ requires the Neumann boundary condition $Z^{zz}=0$.  We also
impose the homogeneous Neumann boundary conditions ${\cal Z}^{ab}=0$ so that
$\partial_z H^a:=\partial_b Z^{ab}+\partial_z^2 \gamma^{az}=0$ requires
$\partial_z^2 \gamma^{az}=0$ at the boundary. Remarkably, subject to the above
conditions, the reduced equation ${\cal E}^{az}=0$ implies $\partial_z^2
\gamma^{az}=0$ at the boundary! Underlying this result is that $S^{az}=0$ at
the boundary due to the local reflection symmetry implied by the above 
conditions. This establishes the maximally dissipative boundary conditions
$H^z=\partial_z H^a=0$ for the constraint propagation equations
(\ref{eq:constrpr}) which ensure that the full Einstein system is satisfied.

In practice, homogeneous boundary conditions do not correspond to a  given
physical problem, e.g. homogeneous Neumann data at the end of a  string lead to
a free endpoint whereas the the endpoint might be undergoing a forced
oscillation requiring inhomogeneous data. This flexibility is supplied within
the maximally dissipative formalism by the ability to extend the homogeneous
boundary condition $Mu=0$ to the inhomogeneous form
$M(x^a)(u-q(x^a))=0$\cite{Friedrich98}. This preserves the well-posedness of
the IBVP for the reduced system with inhomogeneous Neumann data ${\cal
Z}^{zz}=q^{zz}$ and ${\cal Z}^{ab}=q^{ab}$.  For the full system, the gauge
condition $\gamma^{za}=0$ and the boundary constraint $H^z=0$ forces
$q^{zz}=0$. Next, $\partial_z H^a=0$ implies
\begin{equation}
    D_a( {\cal Z}^{ab}/\sqrt{-g^{zz}/g})=0 ,
\label{eq:bconstr}
\end{equation}
where $D_a$  is the connection intrinsic to the boundary. The appearance of the
metric and $D_a$ in Eq.~(\ref{eq:bconstr}) introduces $u$-dependence in the
boundary data so that Secchi's theorems do not apply. However, the theory does
apply to boundary data $\delta q^{ab}$ linearized off a nonlinear solution with
homogeneous data, either exact or generated numerically. Then
Eq.~(\ref{eq:bconstr}) has the form
\begin{equation}
   \partial_b \delta q^{ab}+ F^a_{bc}(x^d) \delta q^{bc}=0,
\label{eq:lbconstr}
\end{equation}
where $F^a_{bc}(x^d)$ is explicitly known via the metric and connection of the
homogeneous solution. The principal part of Eq.~(\ref{eq:lbconstr}) is
identical to the analogous equation in the linearized version of the harmonic
IBVP treated in Ref.~\cite{hbdry}. In terms of the coordinates
$x^a=(t,x^A)=(t,x,y)$ on the boundary, a simple transformation of variables
(see \cite{hbdry}) recasts Eq.~(\ref{eq:lbconstr}) as a symmetric hyperbolic
system of equations for $\phi =\frac{1}{2}\delta_{AB}\delta q^{AB}$ and
$y^A=\delta q^{tA}$. (Here $\delta_{AB}$ is the Kronecker delta.) This system
uniquely determines $\phi$ and $y^A$ in terms of their initial values and the 3
pieces of free boundary data $y^{AB}= \delta q^{AB}+ \delta^{AB}(\delta
q^{tt}-\delta_{CD}\delta q^{CD})$. (This in accord with 
Ref.~\cite{Friedrich98}, although there is no direct correspondence with the
three free pieces of boundary data in~\cite{Friedrich98}). Since only
coordinate conditions have been imposed here, the only restriction on physical
generality is the linearity of the boundary data. 

The IBVP also requires consistency between the Cauchy data and boundary data at
${\cal S}\cup {\cal B}$, which determines the degree of differentiability of
the solution~\cite{secchi2}. As in the string example, consistent homogeneous
Neumann boundary data and Cauchy data imply a virtual reflection symmetry
across the boundary, which is broken in the inhomogeneous case. Although the
IBVP is well-posed for the reduced system and for the constrained system with
boundary data linearized about the homogeneous case, no available theorems
guarantee well-posedness for the constrained inhomogeneous case. In this
respect, the analytic underpinnings are not as general as the Friedrich-Nagy
formulation. Numerical simulations are necessary to shed further light on this
question. The key feature of our formulation is that {\em if a solution exists,
as provided by a convergent numerical simulation, then it necessarily satisfies
the constraints}, since the constraint propagation equation (\ref{eq:constrpr})
is then satisfied with maximally dissipative, homogeneous boundary data. In the
strong field convergence tests described below, exact solutions provide the
Cauchy and boundary data.

In constructing a code to demonstrate these results, we take considerable
liberty with the symmetric hyperbolic formalism. In particular, we use the
second differential order form of the equations based upon the 10 variables
$\gamma^{\alpha\beta}$ rather than the 50 first order variables $u$; we use a
cubic boundary aligned with Cartesian coordinates, although the mathematical
theorems only apply to smooth boundaries; and we replace the gauge condition
$\gamma^{za}=0$ by  $\gamma^{za}=q^a(x^b)\gamma^{zz}$, where $q^a=\partial_z
x^a|_{\cal B}$ is the free Neumann boundary data in the transformation to a
general harmonic coordinate system satisfying $\Box x^a=0$. The harmonic
boundary constraint $H^z=0$ now implies $q^{zz}=-\partial_a(q^a
\gamma^{zz})|_{\cal B}$ and the constraint $\partial_z H^a=0$ again determines
$\phi$ and $y^A$ in terms of the free boundary data $(q^a,y^{AB})$, now through
a symmetric hyperbolic system obtained from adding source terms arising from
$q^a \ne 0$ to the right hand side of Eq.~(\ref{eq:bconstr}). We use the finite
difference techniques described in~\cite{hbdry}, where robust stability and
convergence of a linearized harmonic code was demonstrated. In the linearized
theory, the decoupling of the metric components gives more flexibility in
formulating a well-posed IBVP. The linearized harmonic code could be
consistently implemented with Dirichlet boundary conditions, in which case it
ran stably for $2000$ crossing times even with a piecewise-cubic spherical
boundary cut out of the Cartesian grid. However, we have not found a well-posed
version of the nonlinear theory that avoids Neumann boundary conditions and the
associated numerical complications which we describe below.

\begin{figure}
\begin{psfrags}
\psfrag{Gzz120}[lb]{\small{$|\gamma^{zz}_e|_\infty, A=10^{-1}, 120^3$}}
\psfrag{Gzz80}[lb]{\small{$|\gamma^{zz}_e|_\infty, A=10^{-1}, 80^3$}}
\psfrag{Gzz80S}[lb]{\small{$|\gamma^{zz}_e|_\infty, A=10^{-3}, 80^3$}}
\psfrag{Hn120}[lb]{\small{$|H|_\infty, A=10^{-1}, 120^3$}}
\psfrag{Hn80}[lb]{\small{$|H|_\infty, A=10^{-1}, 80^3$}}
\psfrag{xlabel}[cb]{\small{$t$ (crossing times)}}
\centerline{\epsfxsize=3.1in\epsfbox{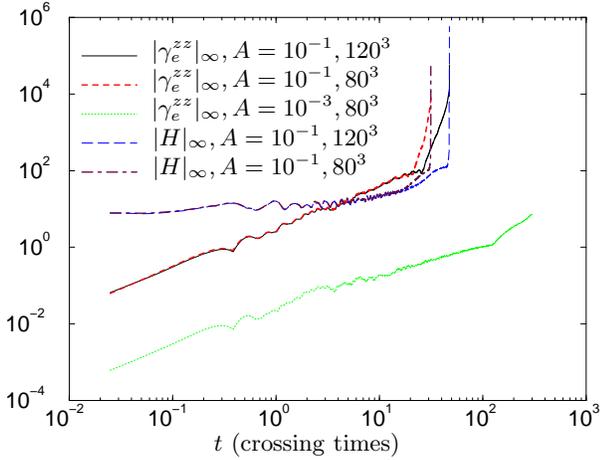}}
\end{psfrags}
\caption{The $L_\infty$ norm of the finite-difference error
$\gamma^{zz}_e = \gamma^{zz}_{ana} - \gamma^{zz}_{num}$,  rescaled  by a
factor of $1/\Delta^2$,  for a gauge-wave. The upper two (mostly overlapping)
curves demonstrate convergence to the analytic solution for a wave with
amplitude $A = 10^{-1}$ 
with gridsizes $80^3$ and
$120^3$. We also plot $|H|_\infty$, the $L_\infty$ norm of $\sqrt
{(H^t)^2+\delta_{ij}H^i  H^j}$, to demonstrate that convergence of the harmonic
constraints is enforced by the boundary conditions. The lower curve represents
evolution of the same gauge-wave with $A = 10^{-3}$ for $300$ crossing times
with gridsize $80^3$.}
\label{fig:gauge.conv}
\end{figure}

\begin{figure}
\begin{psfrags}
\psfrag{xlabel}[cc]{$x$}
\psfrag{ylabel}[cc]{$z$}
\psfrag{title}[c]{$\gamma^{zz}( t = 30 )$}
\centerline{\epsfxsize=3.0in\epsfbox{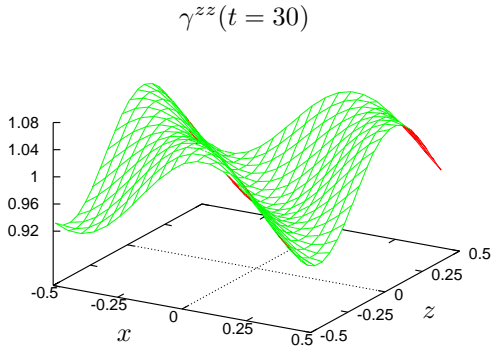}}
\end{psfrags}
\caption{A $y=0$ slice of the metric component $\gamma^{zz}$, evolved for 
$30$ crossing times, amplitude $A = 10^{-1}$, 
with a toroidal boundary in the $(x,y)$ plane.}
\label{fig:2D.full}
\end{figure}

We tested robust stability~\cite{mex} of the nonlinear code by initializing the
evolution with random, constraint violating initial data
$\gamma^{\alpha\beta}=\eta^{\alpha\beta}+\epsilon^{\alpha\beta}$ and by
assigning random boundary data $q(x^a)=\epsilon$ at each point of the cubic
grid boundary, with the $\epsilon$'s random numbers in the range
$(-10^{-10},10^{-10})$. (Although differing from the standard numerical
definition of stability related to convergence, robust stability is
computationally practical for revealing short  wavelength instabilities.) Under
these conditions, the noise in the nonlinear code grows linearly at the same
rate for 2000 crossing times for both the $48^3$ and $72^3$ grids. We tested
convergence in the nonlinear regime using a gauge-wave generated by the
harmonic coordinate transformation $(x,y)=x^B \rightarrow x^B + a^B  \sin\left[
2 \pi \, (\sqrt{3} t + x+y+z)\right]$ acting on the Minkowski metric, with $a^x
= 0.06\, A , \, a^y = 0.04\, A$.  The resulting gauge-wave has amplitude
$||g^{\alpha \beta}-\eta^{\alpha \beta}||_{\infty} \approx A$. We use
periodicity in the $(x,y)$ plane to evolve with smooth toroidal boundaries at
$z=\pm 1/2$. Second-order convergence in the non-linear regime was confirmed
with the amplitude $A = 10^{-1}$. Figures~\ref{fig:gauge.conv} demonstrates the
convergence of the solution and Fig.~\ref{fig:2D.full} shows the absence of
anomalous boundary error. Error arising from the application of Neumann
boundary conditions eventually triggers a nonlinear instability, which occurs
after 30 crossing times with the $120^3$ grid. Runs with the amplitude $A =
10^{-3}$ were carried out on the $80^3$ grid for $300$ crossing times without
encountering the above non-linear instability (see Figure
\ref{fig:gauge.conv}). In the case of a cubic boundary, the nonlinear code
cleanly propagates a physical pulse with amplitude $10^{-7}$ that corresponds
to an exact linearized solution; but, for a gauge-wave of amplitude
$A=10^{-3}$, substantial error arises at the edges and corners due to our
present method of applying Neumann boundary conditions and leads to an
instability after $60$ crossing times. 

The physically proper boundary data for a given problem is a separate and
difficult problem for nonlinear systems. One approach is to supply $q(x^a)$ by
Cauchy-characteristic matching (CCM) in which an interior Cauchy evolution with
cubic boundary is matched to an exterior characteristic evolution on a sequence
of outgoing null cones extending to infinity (for a review see \cite{wliv}). In
simulations of a nonlinear scalar wave with periodic source, CCM was
demonstrated to compute the radiated waveform more efficiently and accurately
than existing artificial boundary conditions on a large but finite
boundary.~\cite{jcp} Previous attempts at CCM in the gravitational case were
plagued by boundary induced instabilities growing on a scale of 10 to 20 grid
crossing times. Although stable behavior of the Cauchy boundary is only {\em a
necessary but not a sufficient} condition for CCM, tests carried out with a
linearized harmonic Cauchy code with a well-posed IBVP matched to a linearized
characteristic code show no instabilities.  

\begin{figure}
\begin{psfrags}
\psfrag{xlabel}[cc]{$x$}
\psfrag{ylabel}[cc]{$y$}
\psfrag{title}[c]{$\gamma^{xy}( t = 0 )$}
\centerline{\epsfxsize=3.0in\epsfbox{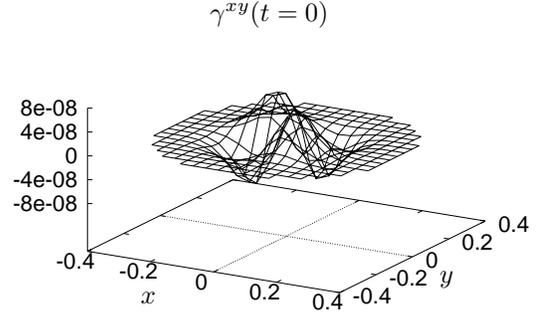}}
\end{psfrags}
\begin{psfrags}
\psfrag{xlabel}[cc]{$x$}
\psfrag{ylabel}[cc]{$y$}
\psfrag{title}[c]{$\gamma^{xy}( t = 1/4 )$}
\centerline{\epsfxsize=3.0in\epsfbox{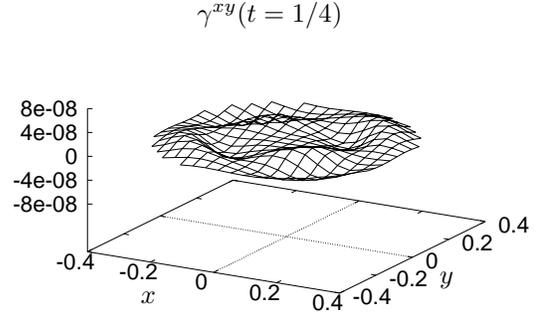}}
\end{psfrags}
\begin{psfrags}
\psfrag{xlabel}[cc]{$x$}
\psfrag{ylabel}[cc]{$y$}
\psfrag{title}[c]{$\gamma^{xy}( t = 1 )$}
\centerline{\epsfxsize=3.0in\epsfbox{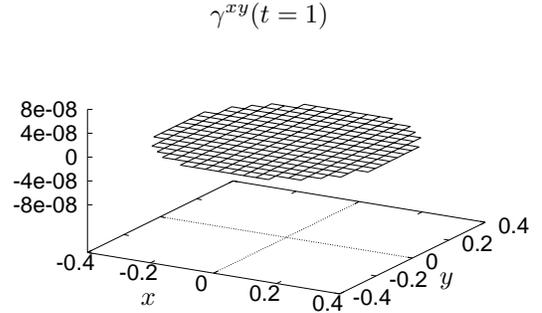}}
\end{psfrags}
\caption{Sequence of $z=0$ slices of the metric component $\gamma^{xy}$,
evolved for one crossing time, with the linear matched Cauchy-characteristic
code. In the last snapshot, the wave has propagated cleanly onto the
characteristic grid.}
\label{fig:2D.ccm}
\end{figure}

In the tests of CCM, the linearized Cauchy code was supplied outer boundary
data $q$ in Sommerfeld form by the exterior characteristic evolution and
boundary data for the characteristic code was supplied on an interior spherical
boundary by the Cauchy evolution. Robust stability for 2000 crossing times on a
Cauchy grid of $45^3$ was confirmed. For a linearized wave pulse,
Figure~\ref{fig:2D.ccm} shows a sequence of profiles of the  metric component
$\gamma^{xy}$ propagating cleanly through the spherical boundary as the wave
pass to the characteristic grid, where it is propagated to infinity. Further
details and tests of CCM and the question of its extension to the nonlinear
theory will be reported elsewhere.

At present, the major limitation in the nonlinear code stems from the
difficulty in handling large values of $\gamma^{tz}$ at the boundary. This is
evidenced by numerical experiments with the manifestly well-posed IBVP
consisting of a scalar wave propagating between smooth toroidal boundaries
according to the flat-space wave equation 
$$\bigg (-\partial_t^2
-2v\partial_t\partial_z+\partial_x^2+\partial_y^2+
(1-v^2)\partial_z^2 \bigg )\Phi=0,$$ 
which arises from the transformation $z\rightarrow z+vt$ on standard inertial
coordinates. The value of $\gamma^{tz}$ represents the velocity of the boundary
relative to observers at rest with respect to the Cauchy slicing. For the flat
space wave equation in second order form, there have apparently been no studies
of numerical algorithms which apply Neumann boundary conditions to such moving
boundaries. In fact, only very recently has there been a thorough treatment of
Neumann boundary conditions for the the flat space wave equation with a
stationary (but curvilinear) boundary~\cite{kreiss2}. This treatment uses
Neumann data to update the field at a boundary point at the current time step
by a one-sided finite difference approximation for the normal derivative. Such
stencils for approximating normal derivatives apply only when the normal
direction is tangent to the Cauchy slicing, i.e. when $g^{tz}|_{\cal B}=0$. The
general case in which $g^{tz}|_{\cal B}\ne0$ requires a more complicated
stencil involving interior points to the future or past of the current time
step. We have developed a new approach which successfully handles this general
case  for the above scalar wave test problem but requires further refinement to
handle boundaries with edges and corners before it can be implemented in the
gravitational code.

We thank H. Friedrich and B. Schmidt for educating us in the intricacy of
the IBVP. The code was parallelized with help from the Cactus development
team of the AEI. The work was supported by NSF grant PHY 9988663.

\newpage

\end{document}